\documentclass[12pt]{iopart}
\usepackage{epsfig}
\usepackage{graphicx}
\usepackage{iopams}
\usepackage{hyperref}
\usepackage{amssymb}
\usepackage{bm,bbm}
\usepackage{color}
\usepackage{threeparttable}
\usepackage[dvipsnames]{xcolor}

\hypersetup{colorlinks,citecolor=blue,filecolor=blue,linkcolor=blue,urlcolor=blue}
\begin{document}

\title{Topological phases of the dimerized Hofstadter butterfly}
\author{Zheng-Wei Zuo$^{1,2}$\footnote{Corresponding author, zuozw@haust.edu.cn}, Wladimir A. Benalcazar$^1$, Yunzhe Liu$^1$ and Chao-Xing Liu$^1$}
\address{$^1$Department of Physics, The Pennsylvania State University, University Park, Pennsylvania 16802, USA}
\address{$^2$School of Physics and Engineering, and Henan Key Laboratory of Photoelectric Energy Storage Materials and Applications, Henan University of Science and Technology, Luoyang 471023, China}

\begin{abstract}
In this work, we study the topological phases of the dimerized square lattice in the presence of an external magnetic field. The dimerization pattern in the lattice's hopping amplitudes can induce a series of bulk energy gap openings in the Hofstadter spectrum at certain fractional fillings, giving rise to various topological phases.
In particular, we show that at $\frac{1}{2}$-filling the topological quadrupole insulator phase with a quadrupole moment quantized to $\frac{e}{2}$ and associated corner-localized mid-gap states exists in certain parameter regime for all magnetic fluxes. At $\frac{1}{4}$ filling, the system can host obstructed atomic limit phases or Chern insulator phases. For those configurations gapped at fillings below $\frac{1}{4}$, the system is in Chern insulator phases of various non-vanishing Chern numbers. Across the phase diagram, both bulk-obstructed and boundary-obstructed topological phase transitions exist in this model.
\end{abstract}

\hspace{4pc}\noindent{Keywords: Topological phases of matter, Topological phase transition,}

\hspace{9pc}\noindent{Topological insulators, Topological materials}

\maketitle

\section{Introduction}

During the past decade, one of the most exciting subjects in condensed matter physics has been the discovery of new topological phases of quantum matter and their identification in material compounds~\cite{ZhangTT19NT,VergnioryMG19NT,TangF19NT}. At present, there is interest in exploring higher-order topological phases (HOTPs)~\cite{Benalcazar17SCI,Benalcazar17PRB,Langbehn17PRL,SongZD17PRL,SchindlerF18SR}, which in $d$-dimensional systems manifest through topologically protected boundary features of dimension less than $d-1$. For example, the hallmark of two-dimensional (2D) second-order topological insulators is the existence of corner fractional charges and/or corner states while their 2D bulk and one-dimensional (1D) edges are both neutral and gapped. Many schemes for realizing HOTPs have been theoretically proposed~\cite{Benalcazar17SCI,Benalcazar17PRB,Langbehn17PRL,SongZD17PRL,SchindlerF18SR,LinM18PRB,EzawaM18PRL,ZhuXY18PRB,KhalafE18PRB,GeierM18PRB,FrancaS18PRB,YanZB18PRL,LiuF19PRL} and the properties of HOTPs have been experimentally demonstrated in photonic, microwave, electronic circuit, and acoustic lattice systems~\cite{noh2018,PetersonCW18NT,Serra-GarciaM18NT,SchindlerF18NTP,ImhofS18NTP,LeeCH18CP,Serra-GarciaM19PRB,XiaoHR19NTM,benalcazar2020}. HOTPs have also been theoretically proposed in several insulators~\cite{SchindlerF18NTP,schindler2019} and superconductors~\cite{benalcazar2014,WangYX18PRB,ZhuXY19PRL,YanZB19PRL,WuXX20PRX}.

At the core of our understanding of first-order topological phases is the concept of a dipole moment in non-magnetic crystalline insulators~\cite{qi2008}. Higher-order topological phases, on the other hand, were initially conceived by finding that crystalline insulators can also host quadrupole and octupole moments~\cite{Benalcazar17SCI}. A quadrupole moment was found in a square lattice model with nearest-neighbor hoppings and a hopping dimerization along both directions to which a $\pi$ flux per plaquette was added. In this paper, we extend this model by considering general fluxes and study their topological phases. We thus turn our attention to the Hofstadter model, which describes charged particles in a lattice in the presence of an external magnetic field~\cite{Hofstadter76PRB}. In this model, the energy spectrum as a function of the magnetic flux has a fractal structure known as Hofstadter Butterfly~\cite{Hofstadter76PRB}. The effect of a magnetic field in electronic systems dates back to von Klitzing et al. They showed that two-dimensional electron gasses in the presence of strong magnetic fields exhibit a robust quantization in their Hall conductance~\cite{vonklitzing1980}. By adding a lattice structure, translation symmetry gets broken down to discrete translation symmetry, giving rise to the Hofstadter model~\cite{Thouless82PRL,Hofstadter76PRB}. Certain variants of the Hofstadter model can host several phenomena, including the existence of edge states due to weak topological phases when hoppings along one direction are dimerized~\cite{LauA15PRL}, the quantum spin Hall effect in a time-reversal-invariant Hofstadter-Hubbard model~\cite{GoldmanN10PRL,CocksD12PRL,UmucalRO17PRL}, and fragile and higher-order topological bands~\cite{LianB20PRB,Herzog-ArbeitmanJ20PRL,WangJ20PRL}. Furthermore, the existence of a fractional Chern insulator phase has been predicted in an interacting Hofstadter model~\cite{WangD13PRL,ScaffidiT14PRB,WuYH15PRB,GersterM17PRB}. Experimentally, the Hofstadter model has been realized in systems of ultracold atoms in optical lattices~\cite{AidelsburgerM13PRL,MiyakeH13PRL,AidelsburgerM15NTP} and Moiré superlattice systems~\cite{HuntB13SCI,DeanC13NT,YuGL14NTP}. More recently, an energetically unbounded and connected multiband  Hofstadter butterfly spectrum in twisted bilayer graphene has been experimentally observed~\cite{LuXB20arXiv}.

In this work, we study a variant of the Hofstadter model in which the hopping amplitudes are dimerized in both directions. The dimerization pattern in the hopping amplitudes opens multiple energy gaps in the system that result in different  topological insulator phases according to their fillings. At $\frac{1}{2}$-filling, the insulating state is either trivial or a quadrupole topological insulator with $\frac{e}{2}$ corner charges at all rational values of the magnetic flux. At $\frac{1}{4}$-filling, the system exists in either Chern insulating phases or obstructed atomic limit phases. At even lower fillings, the number of bulk electrons per elementary unit-cell (defined below) is not enough to construct Wannier representations. In these cases, we only observe Chern insulator phases.

The paper is organized as follows. In Sec.~\ref{sec:Hamiltonian}, we describe the model Hamiltonian and  summarize the main features in the spectrum of the dimerized Hofstadter model (DHM). In the next two sections, we describe the topological phases and their boundary signatures at $\frac{1}{2}$ filling (Sec.~\ref{Sec:half}) and $\frac{1}{4}$ filling (Sec.~\ref{Sec:quarter}). In these two sections, we also describe the phase diagram and the phase transitions in this model, mainly focusing on the case with $\frac{\pi}{2}$ magnetic flux. In Sec.~\ref{Sec:OtherFillings}, we discuss the Chern insulating phases at further fractional fillings. The conclusion is drawn with a brief discussion on the potential experimental feasibility in Sec.~\ref{Sec:Conclusion}.

\section{The Model Hamiltonian and Energy Spectra of DHM}
\label{sec:Hamiltonian}
We start with the description of the Hofstadter model with dimerized hoppings in both directions. The tight-binding Hamiltonian is
\begin{eqnarray}
H=  &  \sum_{m,n}\left[  t_{x}+\left(  -1\right)  ^{m}\delta_{x}\right] e^{i2\pi\phi_{m,n}^{x}}c_{m+1,n}^{\dagger}c_{m,n}\nonumber\\
& + \sum_{m,n}\left[  t_{y}+\left(  -1\right)  ^{n}\delta_{y}\right] e^{i2\pi\phi_{m,n}^{y}}c_{m,n+1}^{\dagger}c_{m,n}+H.c.,
\label{eq:DimHamiltonian}
\end{eqnarray}
where ($m,n$) indicates a lattice site in the ($x,y$) directions, $c^{\dagger}_{m,n}$ and $c_{m,n}$ are fermionic creation and annihilation operators at sites $(m,n)$, $t_{x}$ and $t_{y}$ are the nearest-neighbor average hopping amplitude in $x$ and $y$ directions, respectively, and $\delta_{x}$ and $\delta_{y}$ represent the dimerization amplitudes along $x$ and $y$. $\phi_{m,n}^{x}$ and $\phi_{m,n}^{y}$\ are phase factors that account for the external magnetic field. The magnetic flux per plaquette in units of the magnetic flux quantum is given by $\phi=2\pi(\phi^x_{m,n}+\phi^y_{m+1,n}-\phi^x_{m,n+1}-\phi^y_{m,n})$. From now on, we set $t_{x}=t_{y}=1$. Due to the dimerizations along both $x$ and $y$, each unit cell has 4 sites (see~\ref{GaugeSymmetry}), and we refer to this unit cell as the \emph{elementary unit cell} (EUC) to differentiate it from the \emph{magnetic unit cell} (MUC). We also refer to all fillings with respect to the EUC, e.g., a filling of $\frac{1}{2}$ means the presence of $2$ electrons per EUC.

\begin{figure}[ptbh]
\centering
\includegraphics[scale=1.2]{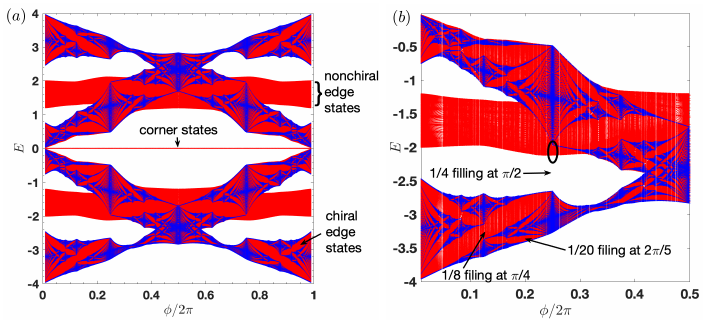}
\caption{(a) Hofstadter butterfly with $(\delta_x,\delta_y)=(0.6,0.6)$. Blue (red) regions are energy bands with periodic (open) boundaries along all directions. The red line at $E=0$ is four-fold degenerate and has corner-localized eigenstates. (b) Zoom-in of the lower left region of the butterfly in (a). The gaps above $\frac{1}{4}$ filling at flux $\pi/2$, $\frac{1}{8}$ filling at flux $\pi/4$, and $\frac{1}{20}$ filling at a flux of $2\pi/5$ are labelled.}
\label{fig:Topology}
\end{figure}

Here we focus on the cases with rational magnetic fluxes $\phi= 2\pi p/q$, with $p, q$ being mutually prime integers. In the absence of the dimerization, the system reduces to the original Hofstadter model, and the MUC contains $q$ lattice sites, so that the bulk spectrum consists of $q$ bands~\cite{Hofstadter76PRB}. When $q$ is even, the bulk spectrum is gapless with $q$ bulk Dirac points at $\frac{1}{2}$ filling~\cite{Wen89NPB,KohmotoM89PRB}. In the presence of hopping dimerization, the sets of gauges that give rise to the magnetic flux depend on the value of $q$ and the dimerization pattern (see the details in~\ref{GaugeSymmetry}). Figure~\ref{fig:Topology}a shows the Hofstadter butterfly for a system with $\delta_x=\delta_y=0.6$. The blue color corresponds to the energy spectra of a system with periodic boundary conditions (PBC), while the red color corresponds to that of a system with open boundary conditions (OBC) \emph{along both $x$ and $y$ directions}. The most salient feature is that several energy bands with OBC are present within the bulk energy gaps under PBC at various fillings, suggesting the existence of in-gap boundary states (corner or edge localized). We will show that these boundary states result from the bulk-boundary correspondence of different topological phases. Here we will first summarize the main features of the energy spectrum and the corresponding topological phases at several different fillings, and the detailed studies will be presented in the subsequent sections:
(a) At the half-filling, zero energy states appear within the bulk energy gap for all values of $\phi$ under OBC (red lines in Fig.~\ref{fig:Topology}a). These zero-energy states are fourfold degenerate, and each state is located at a corner of the lattice. The system is in the topological quadrupole insulator (TQI) phase ~\cite{Benalcazar17SCI} with the quantized quadrupole moment operator $Q_{xy}^{p}=\frac{1}{2}$~\cite{Wheeler18arXiv,KangBM18arXiv} protected by chiral symmetry~\cite{chang2020}~\footnote{This TQI phase at general values of the flux has also been recently studied in Ref.~\cite{OtakiK19arXiv}}. (b) At $\frac{1}{4}$ filling, we zoom in the lower left corner of the energy spectrum (Fig. \ref{fig:Topology}b). There, one can see an in-gap red band within the gap of the bulk states. As discussed in Sec. \ref{Sec:quarter}, these in-gap states are nonchiral edge states and originate from obstructed atomic limit (OAL) phases protected by $C_2$ symmetry at values of the flux. However, depending on the dimerizations' values, Chern insulating phases can also exist.
(c) For fillings below $\frac{1}{4}$, we find a Chern insulator phase at $\frac{1}{8}$ filling for a flux of $\pi/4$, and at a filling of $\frac{1}{20}$ for a flux of $2\pi/5$, as marked in Fig.~\ref{fig:Topology}b. In Sec. \ref{Sec:OtherFillings}, we argue that this is a generic feature of any gapped state with a filling less than $\frac{1}{4}$ since in such case, there is less than one electron per EUC, making a Wannier function description is impossible.

\section{Phases at $\frac{1}{2}$ filling}\label{Sec:half}
We first discuss the phases at $\phi=\frac{\pi}{2}$, then at other fluxes, and finally, we discuss the phase transitions within the phase diagram.

\subsection{TQI phases at $\phi=\frac{\pi}{2}$}
\begin{figure}[ptbh]
\centering
\includegraphics[scale=0.9]{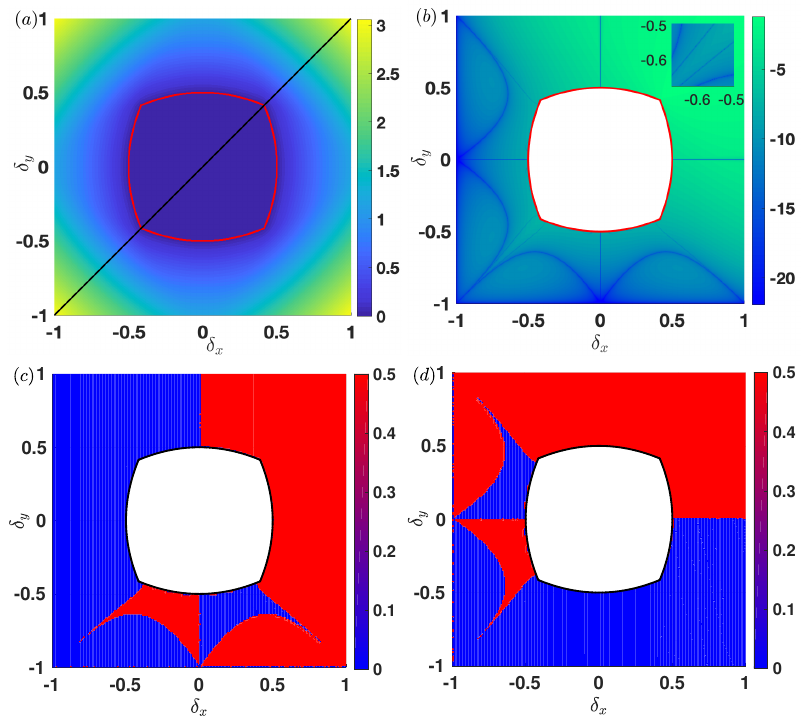}
\caption{Energy gap and Wannier gaps at $\frac{1}{2}$ filling for flux $\phi=\frac{\pi}{2}$. (a) Energy gap. The region inside the red line is gapless and consequently its Wannier bands cannot be defined. Along the black diagonal line, the system obeys $C_4$ symmetry. (b) Wannier band gap in logarithmic scale. The inset shows the Wannier gap closing curves around the parameter point $(\delta_{x},\delta_{y})=(-0.6,-0.6)$. (c,d) Wannier-sector polarization $p_{y}^{\nu_x}$ (c), and $p_{x}^{\nu_y}$ (d).}
\label{fig:Gap}
\end{figure}

\begin{figure}[tbh]
\centering
\includegraphics[scale=1.2]{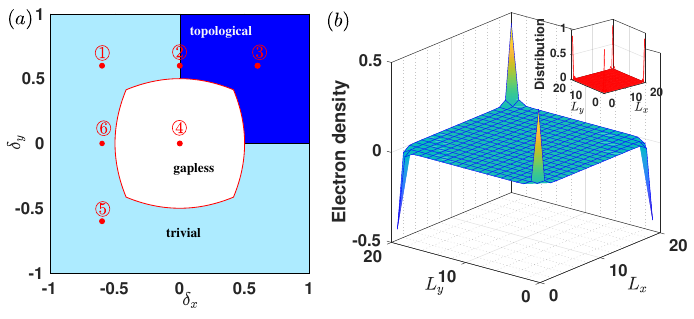}
\caption{(a) Phase diagram at $\frac{1}{2}$-filling for flux $\phi=\frac{\pi}{2}$ as a function of the dimerization variables $\delta_x$ and $\delta_y$. The points marked by numbers $\textcircled{1}$-$\textcircled{6}$ have the Hofstadter butterflies shown in Fig.~\ref{fig:Butterfly}. The values of the parameters in these six points are $(\delta_x,\delta_y)=(-0.6,0.6)$, $(0,0.6)$,$(0.6,0.6)$,$(0,0)$,$(-0.6,-0.6)$, and $(-0.6, 0)$. (b) Charge density for the quadrupole phase at $\frac{1}{2}$-filling for flux $\phi=\frac{\pi}{2}$ (the inset shows the zero-energy corner states) with $(\delta_x,\delta_y)=(0.6,0.6)$. The system size is $L_x=L_y=20$ EUCs.
}
\label{fig:PhaseDiagramHalf}
\end{figure}

At $\frac{1}{2}$ filling, the quadrupole moment is quantized, and the corners carry fractionally quantized charges of $\frac{e}{2}$. When $\phi=\pi$, this model is identical to the minimal model initially proposed for a quadrupole topological insulator~\cite{Benalcazar17SCI}. At other values of $\phi$, the gap at $\frac{1}{2}$ filling remains open for strong amplitudes of dimerization (how strong the dimerization must be depends on the flux, see below) as shown in Fig.~\ref{fig:Topology}a, and thus the TQI phase is preserved due to the conservation of chiral symmetry for all values of $\phi$. We take the $\phi=\frac{\pi}{2}$ case (for which $p/q=\frac{1}{4}$) as an example. Unlike the case with $\pi$ flux, which only has gapless bulk bands at $\frac{1}{2}$ filling at the point $(\delta_x,\delta_y)=(0,0)$~\cite{Benalcazar17PRB}, a larger gapless region exists for $\frac{\pi}{2}$ flux, enclosed by the red curve in Fig.~\ref{fig:Gap}a. The Wannier gaps and Wannier-sector polarization \cite{Benalcazar17SCI,Benalcazar17PRB} $p_{y}^{\nu_x}$, and $p_{x}^{\nu_y}$ are computed in Fig.~\ref{fig:Gap}b-d, from which we construct the complete phase diagram at $\frac{1}{2}$ filling for the flux $\frac{\pi}{2}$, as shown in Fig.\ref{fig:PhaseDiagramHalf}a.

In the region enclosed by the red curve in Fig.~\ref{fig:Gap}a, Wannier bands cannot be defined because the system is gapless. Outside of the gapless region, Fig.~\ref{fig:Gap}b shows the values of the Wannier gaps, which close at several dark blues lines. Due to the existence of reflection symmetries at $\phi=\frac{\pi}{2}$, the Wannier sector polarizations $p_{y}^{v_{x}}$ and $p_{x}^{v_{y}}$~\cite{Benalcazar17SCI,Benalcazar17PRB} are quantized to $0$ or  $\frac{1}{2}$, and their values can only change by closing the Wannier gaps. The values of the Wannier sector polarizations are shown in Fig.~\ref{fig:Gap}c,d, where the blue and red colors correspond to 0 and $0.5$, respectively. Indeed, the jumps in the values of $p_{y}^{\nu_{x}}$ and $p_{x}^{\nu_{y}}$ in Fig.\ref{fig:Gap}c,d coincide with the Wannier gap closing lines in Fig.~\ref{fig:Gap}a. The TQI phase corresponds to the region in the phase diagram where both Wannier sector polarizations take the nontrivial value of $0.5$. With the information extracted from Fig.~\ref{fig:Gap}, the phase diagram at $\phi=\frac{\pi}{2}$ is built, as shown in Fig.\ref{fig:PhaseDiagramHalf}a. In the blue and light blue regions, the system is in the TQI phase and the trivial phase, respectively. In the middle white region, the system is gapless at $\frac{1}{2}$ filling.

Figure~\ref{fig:PhaseDiagramHalf}b shows the electronic charge density at $\frac{1}{2}$ filling relative to the background ionic charge of $+2e$ per EUC for a value of $(\delta_x,\delta_y)=(0.6,0.6)$. While the bulk has an overall vanishing charge, there are corner-localized charges of $\pm \frac{e}{2}$. Along with these charges, four degenerate states at zero energy localize at four corners (one state per corner, corresponding to the zero-energy red lines at $E=0$ in Fig.~\ref{fig:Topology}a), and are protected by chiral symmetry~\cite{chang2020} [in order to fix the sign of the quadrupole moment and its corner charges, we add infinitesimal onsite potentials $\mu$ ($-\mu$) on the lattice sites $A$ and $D$ ($B$ and $C$), as labeled in Fig.\ref{app:Pi/2}, which break chiral symmetry, both reflections, and $C_4$ symmetries, but preserves $C_2$/inversion symmetry, which is necessary to fix the bulk polarization to identically zero, see ~\ref{GaugeSymmetry}].

\subsection{TQI phases at other values of flux}
The quantization of the Wannier sector polarizations $p_{y}^{v_{x}}$ and $p_{x}^{v_{y}}$, as well as the quadrupole moment $Q_{xy}=2p_{x}^{\nu_y}p_{y}^{\nu_x}$~\cite{Benalcazar17SCI,Benalcazar17PRB} requires $C_4$ symmetry or reflection symmetries. However, $C_4$ symmetry and reflection symmetries only exist at the specific flux values $\phi=\pi$ or $\frac{\pi}{2}$, while at the generic $\phi$ values, the DHM only possesses $C_2$ rotation symmetry. Both $C_2$ symmetry and chiral symmetry pin the bulk polarization to exactly zero, enabling the quadrupole moment to be well-defined. Indeed, chiral symmetry alone is sufficient to quantize the quadrupole moment to $Q_{xy}=\frac{1}{2}$~\cite{chang2020}. To evaluate $Q_{xy}$, we use the operator $Q_{x y}^{p}= \frac{1}{2\pi} \mathrm{Im} \log \left\langle \Psi_{0}\left| e^{\frac{2 \pi i}{L _x L_y} \sum_{x,y} \hat{q}_{x y}} \right| \Psi_{0}\right\rangle$, where $\hat{q}_{xy}=\hat{x}\hat{y}$ is the quadrupole moment operator, and $L_x$, $L_y$ are the number of MUCs along the $x$ and $y$ directions, respectively \cite{Wheeler18arXiv,KangBM18arXiv}.  $\Psi_{0}$ is the many-body ground state.
Indeed, the quadrupole moment remains quantized to $Q_{xy} ^{p}=0.5$ at all flux values. The TQI phases at general fluxes have recently been identified in Refs.~\cite{OtakiK19arXiv,AsagaK20PRB} using the complementary framework of entanglement polarization.

\subsection{Topological Phase transitions}
\begin{figure}[ptbh]
\centering
\includegraphics[scale=1.1]{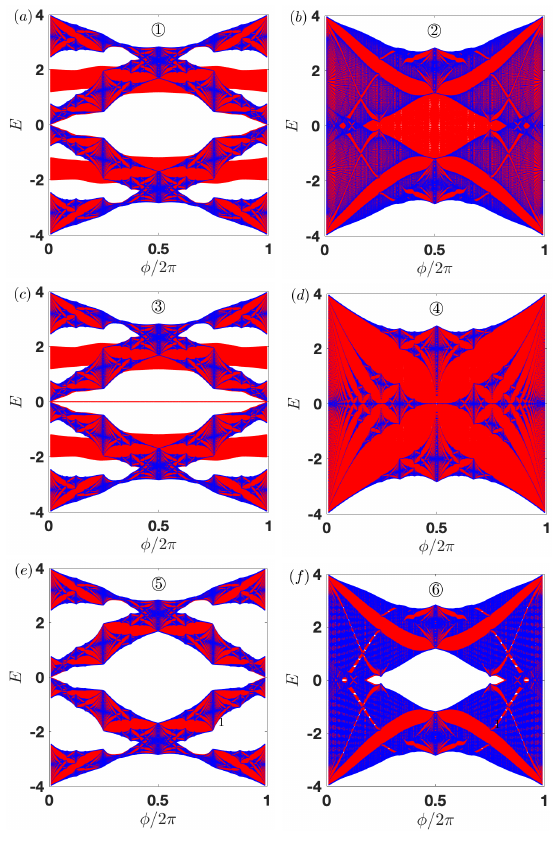}
\caption{(a-f) Hofstadter butterflies for dimerization values $(\delta_x,\delta_y)=(-0.6,0.6)$, $(0,0.6)$,$(0.6,0.6)$,$(0,0)$,$(-0.6,-0.6)$, and $(-0.6, 0)$, respectively. Blue (red) spectra are obtained with periodic (open) boundaries along both directions.}
\label{fig:Butterfly}
\end{figure}

After determining the structure of the phase diagram at $\phi=\frac{\pi}{2}$ and the TQI phase for different values of flux $\phi$, we explore the possible phase transitions. In particular, we distinguish two types of phase transitions in the DHM: (i) bulk-obstructed, and (ii) boundary-obstructed~\cite{Benalcazar17PRB,WuXX20PRX,BOTP}. We do so for all values of $\phi$ by plotting the energy spectrum with open (red) and closed (blue) boundaries in Fig.~\ref{fig:Butterfly}. These plots are done at certain specific points $(\delta_x,\delta_y)$ indicated in Fig.~\ref{fig:PhaseDiagramHalf}a. A phase transition that passes through the central part of the phase diagram, $(\delta_x,\delta_y)=(0,0)$, involves a bulk phase transition for all values of the flux $\phi$, as shown by Fig.~\ref{fig:Butterfly}d (point $\raisebox{.5pt}{\textcircled{\raisebox{-.9pt} {4}}}$ in Fig.\ref{fig:PhaseDiagramHalf}a), in which the blue energy spectrum closes the gap at $\frac{1}{2}$ filling. This is expected as the point $(\delta_x,\delta_y)=(0,0)$ corresponds to the original Hofstadter butterfly model which is gapless. From the point of view of bulk topology, we may first consider the phase transition along the diagonal line  $\delta_x=\delta_y$, along which $C_4$ symmetry is preserved for $\phi=\frac{\pi}{2}$. This line connects the trivial and topological quadrupole phases indicated by points $\raisebox{.5pt}{\textcircled{\raisebox{-.9pt} {3}}}$ and $\raisebox{.5pt}{\textcircled{\raisebox{-.9pt} {5}}}$ in Fig.\ref{fig:PhaseDiagramHalf}a with corresponding energy spectra extended to all values of flux shown in Fig.~\ref{fig:Butterfly}(c) and (e). This bulk-obstructed topological phase transition is similar to that at flux $\phi=\pi$. However, there is one difference; for flux $\phi=\pi$, the transition only occurs at $(\delta_x,\delta_y)=(0,0)$, while for $\phi=\frac{\pi}{2}$, there is a larger gapless region, as shown in Fig.~\ref{fig:PhaseDiagramHalf}b. The exact range of this gapless region is determined by Eq.~\ref{gap-closing} in Appendix.

In the absence of $C_4$ symmetry (e.g., by making $\delta_x \neq \delta_y$), a phase transition can occur which does not need to close the bulk gap but only the 1D edge gap~\cite{Benalcazar17PRB,BOTP}. Here we consider the transition along the line connecting the points  $\raisebox{.5pt}{\textcircled{\raisebox{-.9pt} {3}}}$-$\raisebox{.5pt}{\textcircled{\raisebox{-.9pt} {2}}}$-$\raisebox{.5pt}{\textcircled{\raisebox{-.9pt} {1}}}$-$\raisebox{.5pt}{\textcircled{\raisebox{-.9pt} {6}}}$-$\raisebox{.5pt}{\textcircled{\raisebox{-.9pt} {5}}}$ in Fig.\ref{fig:PhaseDiagramHalf}a. Figure~\ref{fig:Butterfly}b and \ref{fig:Butterfly}f show the Butterfly spectra at $(\delta_x,\delta_y)=(0,0.6)$ and at $(\delta_x,\delta_y)=(-0.6,0)$, respectively. While the bulk energy spectrum (blue spectra in Fig.\ref{fig:Butterfly}b) does not always close its gap, the edge spectrum (red spectra in Fig.\ref{fig:Butterfly}b) does. Closing only the edge spectrum corresponds to a boundary-obstructed topological phase transition~\cite{Benalcazar17PRB,WuXX20PRX,BOTP}. The boundary obstruction is encoded in the topology of the Wannier bands~\cite{klich2011}, so that phase transitions that close only the edge gap are manifested in the bulk by the closing of their Wannier gaps (Fig.~\ref{fig:Gap}b,c,d at $\phi=\frac{\pi}{2}$). Notice that for fluxes close to $\phi=0$, the bulk also closes its gaps at $\frac{1}{2}$ filling. In general, the quantized quadrupole phase can either be boundary obstructed or bulk obstructed. As seen in Fig.~\ref{fig:Butterfly}b, in the presence of only chiral symmetry, at least a boundary obstruction is guaranteed. When the hopping amplitudes are dimerized only in one direction (Fig.~\ref{fig:Butterfly}b and \ref{fig:Butterfly}f), there are weak-topological edge states (i.e., due to a quantization of the Berry phase along the dimerized direction) depending on the dimerization strength at half filling (Fig.~\ref{fig:Butterfly}b), also shown in Ref.~\cite{LauA15PRL}.

\section{Phases at $\frac{1}{4}$ filling}
\label{Sec:quarter}

\begin{figure}[ptbh]
\centering
\includegraphics[scale=1.1]{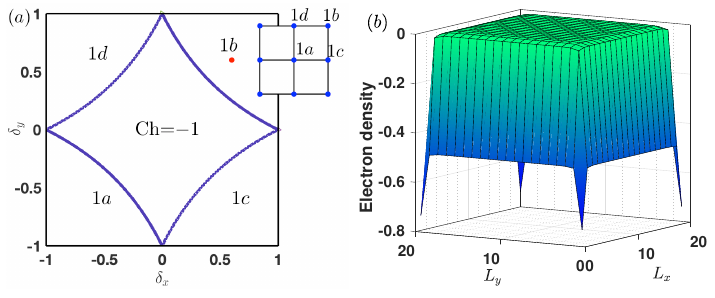}
\caption{(a) Phase diagram at $\frac{1}{4}$ filling and $\phi=\frac{\pi}{2}$ as a function of the dimerization variables $\delta_x$ and $\delta_y$. In the center region, the system is Chern insulator. The inset shows the different maximal Wyckoff position of the EUC protected by $C_2$ symmetry. The $1a$, $1b$, $1c$, and $1d$ stand for the Wannier center pinned to the corresponding Wyckoff position of the EUC. (b) Charge density after filling the lowest $(L_x-1)(L_y-1)$ states (close to $\frac{1}{4}$-filling) under OBC for flux $\phi=\frac{\pi}{2}$ with $(\delta_x,\delta_y)=(0.6,0.6)$ [red dot in (a)].  The system size is $L_x=L_y=20$ EUCs.}
\label{fig:PhaseDiagramFourth}
\end{figure}

In addition to the gap opening at $\frac{1}{2}$ filling, the dimerization of the hopping terms in Eq.~\ref{eq:DimHamiltonian} also opens gaps at $\frac{1}{4}$ filling for most values of the flux. For example, at $\frac{1}{4}$ filling and for the values $(\delta_x,\delta_y)=(0.6,0.6)$, energy gaps exist for magnetic fluxes within the ranges $(0,0.8\pi)$ and $(1.2\pi,2\pi)$ (Fig.~\ref{fig:Topology}b). At $\frac{1}{4}$ filling and for a flux of $\phi=\frac{\pi}{2}$, each EUC has one electron which is either localized to a maximal Myckoff position of the EUC by $C_2$ symmetry, or is delocalized. Accordingly, we find that the insulator is in a trivial insulator phase, an OAL phase~\cite{bradlyn2017, benalcazar2019fillinganomaly} or a Chern insulator phase. Figure~\ref{fig:PhaseDiagramFourth}a shows the phase diagram at $\frac{1}{4}$ filling with $\phi=\frac{\pi}{2}$, which is separated into five phases by the bulk gap closing lines depicted by the blue curves. In the central region, the system is in a Chern insulator phase with Chern number $C=-1$. The other four regions belong to different atomic limits characterized by Wannier centers localized at maximal Wyckoff positions of the EUC (see inset in Fig.~\ref{fig:PhaseDiagramFourth}a for the maximal Wyckoff positions within the $C_2$ symmetric EUC).

In the lower left region in Fig.~\ref{fig:PhaseDiagramFourth}a, labelled $1a$, the model is in a trivial atomic limit, with the Wannier center of the electron pinned by $C_2$ symmetry (also inversion symmetry) to the $1a$ Wyckoff position of the EUC. The insulator in this phase has a uniform charge density in the bulk, edges, and corners.

In the top right region, labelled $1b$ in Fig.~\ref{fig:PhaseDiagramFourth}a, the Wannier centers are pinned by $C_2$ symmetry to the corner of the EUC. This manifests in a bulk dipole moment of ${\bf P}=(\frac{e}{2},\frac{e}{2})$. When the Fermi level is such that electrons occupy only bulk states, the dipole moments result in a charge deficit per unit length of $\frac{e}{2}$ at edges; additionally, there is a charge deficit at each corner of $\frac{e}{4}$, signaling that this configuration could be used in conjunction with another phase that trivializes the bulk dipole moment to generate a second-order topological phase with corner-induced filling anomaly~\cite{benalcazar2019fillinganomaly}. This charge deficit is depicted in Figure~\ref{fig:PhaseDiagramFourth}b, which shows the electron density upon filling of the lowest $(L_x-1)(L_y-1)$ energy levels for system size $L_x=L_y=20$ EUCs under OBC (a filling of $L_xL_y$ states corresponds to exact $\frac{1}{4}$ filling) when the flux is $\phi=\frac{\pi}{2}$ with $(\delta_x,\delta_y)=(0.6,0.6)$, corresponding to the $1b$ OAL phase.
Above that Fermi level, there are edge and corner states [red bands between the first and second bulk bands and in between the third and fourth bulk bands, as (partially) enclosed by the black ellipse in Fig.~\ref{fig:Topology}b].

In the top left region, labelled $1d$ in Fig.~\ref{fig:PhaseDiagramFourth}a, and bottom right region, labelled $1c$, the Wannier centers of the electrons are pinned by $C_2$ symmetry to the maximal positions $1d$ and $1c$, leading to polarizations ${\bf P}=(\frac{e}{2},0)$, and $(0,\frac{e}{2})$, respectively.

All these distinct atomic limits can be determined from the $C_2$ symmetry irreducible representations that the occupied eigenstates adopt at the high-symmetry points of the Brillouin zone, as shown in Table ~\ref{tab:C2Eigenvalues}. Since $C_2$ symmetry is preserved at all values of the flux, similar symmetry indicators can diagnose the topological phases at $\frac{1}{4}$ filling at other fluxes (see ~\ref{OtherFlux} for the cases with flux $\frac{\pi}{4}$ and $\frac{2\pi}{3}$).

\begin{table}
\caption{Eigenvalues of the $C_2$ symmetry operator (see Eq.~\ref{C2Operator} in ~\ref{GaugeSymmetry}), projected into the lowest occupied band at the high-symmetry points of the Brillouin zone ${\bf \Gamma}=(0,0)$, ${\bf X}=(\pi,0)$, ${\bf Y}=(\pi,0)$, and ${\bf M}=(\pi,\pi)$ for flux $\frac{\pi}{2}$ and the corresponding topological phases for different dimerization values $(\delta_x , \delta_y)$.}
\begin{center}
\begin{tabular}[c]{lrrrrc}
\hline
$(\delta_{x}$,$\delta_{y})$ & ${\bf \Gamma}$ & ${\bf X}$ & ${\bf Y}$ & ${\bf M}$ & Wannier center\\ \hline
(0,0) & -1 & -1 & -1 & 1 & none, Chern insulator phase with C=-1 \\
(0.5,0.5) & 1 & -1 & -1 & 1 & 1b \\
(-0.5,0.5) & -1 & -1 & 1 & 1 & 1d\\
(-0.5,-0.5) & -1 & -1 & -1 & -1 & 1a\\
(0.5,-0.5) & -1 & 1 & -1 & 1 & 1c\\ \hline
\end{tabular}
\end{center}
\label{tab:C2Eigenvalues}
\end{table}

\section{Phases at other fractional fillings}
\label{Sec:OtherFillings}
Figure~\ref{fig:Topology}b zooms in the lower left region of the butterfly at $(\delta_x,\delta_y)=(0.6,0.6)$, and illustrates the bulk energy gaps and in-gap boundary states at $\frac{1}{4}$ filling, $\frac{1}{8}$ filling, and at other fractional fillings.
For fillings below $\frac{1}{4}$, we find that the insulators are in integer quantum Hall phases with nonzero Chern numbers. For example, we numerically obtain a Chern number of $C=1$ at $\frac{1}{8}$ filling for a flux of $\frac{\pi}{4}$, and $C=-1$ at $\frac{1}{20}$ filling for a flux of $\frac{2\pi}{5}$. With open boundaries, chiral edge states traverse the bulk energy gaps, as evidenced in Fig.~\ref{fig:Topology}b by the fact that bulk energy gaps are filled with (red) edge states.

These Chern insulator phases are incompatible with a Wannier representation, which is consistent with the fact that at any filling below $\frac{1}{4}$, there is less than one electron per EUC. Although the magnetic field generally enlarges the EUC to a MUC (see~\ref{GaugeSymmetry}), all EUCs are threaded by the same magnetic flux. Therefore, they are \emph{physically} equivalent. As such, it is inconceivable that electrons would distribute inequivalently among the EUCs since our problem here is for non-interaction electrons. Indeed, as shown in \ref{Sec:Proof}, by combining the Diophantine equation and the streda's formula, we can prove that the system is in either a metallic phase or a Chern insulator phase with non-zero Chern number for the filling smaller than $1/4$.


\section{Discussion and conclusion}
\label{Sec:Conclusion}
The dimerized Hofstadter model can give rise to topological phases of three different types. By adjusting the dimerization strength, the system can be in a trivial or topological quadrupole phase at $\frac{1}{2}$ filling at all non-vanishing rational values of the magnetic flux. At $\frac{1}{4}$ filling, all possible OAL phases, along with a Chern insulator phase, can be generated. Finally, at fillings lower than $\frac{1}{4}$, only Chern insulator phases arise. The dimerized Hofstadter model can be realized by a system of ultra-cold atoms in an optical lattice using laser-assisted tunneling and a potential energy gradient provided by magnetic fields, as shown by Refs.~\cite{AidelsburgerM13PRL,MiyakeH13PRL}. For the dimerization in two directions, two orthogonal optical waves can be used in each direction at twice the wavelength of the optical lattice. The tunable flux on the plaquette can be produced by the magnetic field gradient and two tunable frequency Raman beams, which induces a position-dependent phase factor hopping. Also, microwave cavity arrays~\cite{AndersonBM16PRX} and LC circuit networks~\cite{NingyuanJ15PRX} can simulate the DHM. The variety of topological phases within the DHM makes this platform attractive to study several phenomena, including the effect of the disorder, the role of irrational/nonuniform magnetic flux, and electron-electron interactions in its topological phases.

\ack Z.W.Z. is grateful to the National Science Foundation of China (Grants No. 12074101, and No. 11604081) and China Scholarship Council for financial support. Z.Z.W. is also sponsored by Natural Science Foundation of Henan (Grant No. 212300410040). C.X.L. acknowledges the support of the Office of Naval Research (Grant No. N00014-18-1-2793) and Kaufman New Initiative research grant KA2018-98553 of the Pittsburgh Foundation.  W. A. B. thanks the support of the Eberly Posdoctoral Fellowship at the Pennsylvania State University.

\section*{References}

\providecommand{\newblock}{}

\appendix

\section{Gauges for general flux and symmetries of Hamiltonian for flux $\pi/2$}
\label{GaugeSymmetry}
\begin{figure}[ptbh]
\centering
\includegraphics[scale=1]{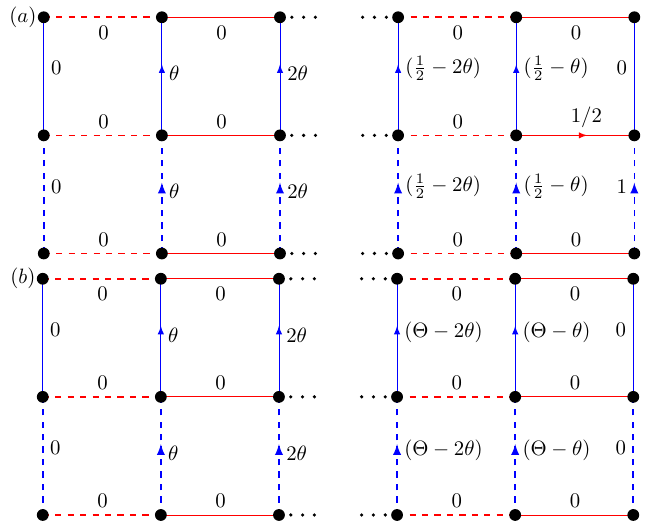}
\caption{The gauge choices for the different rational magnetic flux $\phi= 2\pi p/q=2\pi \theta$ ($p, q$ being mutually prime integers) $(a)$ the gauge for $q=4n$ ($n$ integer number), $(b)$ the gauge for $q=4n+2$ or $q$ odd number. For $q=4n+2$, the values  $\Theta=1$. For $q$ odd number, $\Theta=2$.}
\label{fig:Gauge}
\end{figure}

In this Appendix, we give further details on the determination of the gauge choices that implement a general magnetic flux in the Hofstadter model of Eq.~\ref{eq:DimHamiltonian}. Generally, there are different gauge choices for a specific flux. We choose a gauge illustrated in Fig. ~\ref{fig:Gauge}. The numbers in Fig.~\ref{fig:Gauge} indicate the value of $\theta$ in the Peierls phase factor $e^{i 2\pi \theta}$. For instance, a value of $\frac{1}{4}$ corresponds to a phase factor of $i$. Thus, we can easily write down the Hamiltonians for different fluxes and get these Hofstadter butterflies, as shown in Fig. ~\ref{fig:Butterfly}.

In general, the DHM has $C_2$ symmetry for all values of the flux. At particular values of the flux, additional crystalline symmetries can be present. For example, at values at which $\phi = -\phi $ mod $2\pi$, reflection symmetries also exist. Finally, the choice of gauge may difficult the general existence of $C_4$ symmetry, as the MUC is in general rectangular.

\begin{figure}[ptbh]
\centering
\includegraphics[scale=1]{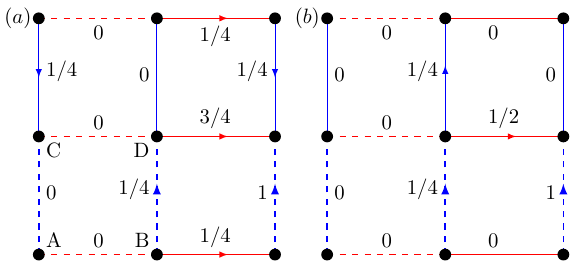}
\caption{Two gauge choices for the implementation of a flux of $\frac{\pi}{2}$ $(p/q=\frac{1}{4})$.}
\label{app:Pi/2}
\end{figure}

To look into the symmetries of the DHM, let us take flux $\pi/2$ case as an example. Two gauge choices for the same flux are shown in Fig.~\ref{app:Pi/2} [the choice in Fig.~\ref{app:Pi/2}b is the same as that in Fig.~\ref{fig:Gauge}a]. By introducing the four-component operator $C_k^{\dagger}=(c_{k A}^{\dagger}, c_{k B}^{\dagger}, c_{kC}^{\dagger}, c_{kD}^{\dagger})$, we can express the Bloch Hamiltonian in the momentum space for the gauge choice in Fig.~\ref{app:Pi/2}a as
\begin{equation}
h=\left(
\begin{array}{cccc}
{0} & {\alpha} & {\beta} & {0} \\
{\alpha^*} & {0} & {0} & {\gamma} \\
{\beta^*} & {0} & {0} & {\lambda} \\
{0} & {\gamma^*} & {\lambda^*} & {0}\end{array}
\right),
\label{Bloch Hamiltonian4}
\end{equation}
where $\alpha=(t_{x}-\delta_{x})+(t_{x}+\delta_{x})e^{i(k_x+\frac{\pi}{2})}$, $\beta=(t_{y}-\delta_{y})+(t_{y}+\delta_{y})e^{i(k_y-\frac{\pi}{2})}$, $\gamma=(t_{y}-\delta_{y})e^{-i\frac{\pi}{2}}+(t_{y}+\delta_{y})e^{ik_y}$, and $\lambda=(t_{x}-\delta_{x})+(t_{x}+\delta_{x})e^{i(k_x+3\frac{\pi}{2})}$. For the gauge choice in Fig.~\ref{app:Pi/2}b, the model Hamiltonian can also be written down in a similar manner.

Now, we discuss the reflection, $C_2$ rotation, and chiral (sublattice) symmetries of the system Hamiltonian in Eq.~\ref{Bloch Hamiltonian4} for flux $\pi/2$. The Hamiltonian in Eq.~\ref{Bloch Hamiltonian4} has chiral symmetry, defined by
\begin{equation}
\hat{\Pi} h_{k} \hat{\Pi}^{-1}=-h_{k}, \quad
\hat{\Pi}=\left(
\begin{array}
[c]{cccc}%
1 & 0 & 0 & 0\\
0 & -1 & 0 & 0\\
0 & 0 & -1 & 0\\
0 & 0 & 0 & 1
\end{array}
\right).
\end{equation}

Under reflection symmetry, the magnetic flux generally changes sign. The reflection symmetries $M_x$, $M_y$ along the $x$ and $y$ directions, respectively, can be written as
\begin{equation}
\begin{array}{l}{M_{x} h\left(k_{x}, k_{y}, \phi\right) M_{x}^{-1}=h\left(-k_{x}, k_{y},-\phi\right)} \\
 {M_{y} h\left(k_{x}, k_{y}, \phi\right) M_{y}^{-1}=h\left(k_{x},-k_{y},-\phi\right)},\end{array}
\end{equation}
where  $M_x$ and $M_y$ are
\begin{equation}
M_{x}=\left(
\begin{array}{cccc}{0} & {1} & {0} & {0} \\
 1 & {0} & {0} & {0} \\
  0 & {0} & {0} & {i} \\
   0 & {0} & {i} & {0}
   \end{array}\right), \quad
M_{y}=\left(
\begin{array}{cccc}{0} & {0} & {1} & {0} \\
{0} & {0} & {0} & {1} \\
 1 & {0} & {0} & {0} \\
  0 & {1} & {0} & {0}\end{array}
  \right).
\end{equation}
Due to the $C_2$ inversion symmetry, we have
\begin{equation}
{C_{2} h\left(k_{x}, k_{y}\right) C_{2}^{-1}=h\left(-k_{x}, -k_{y}\right)},
\quad
C_{2}=\left(
\begin{array}
[c]{cccc}
0 & 0 & 0 &e^{i\frac{5\pi}{4}}\\
0 & 0 &e^{i\frac{5\pi}{4}} & 0\\
0 & e^{i\frac{3\pi}{4}} & 0 & 0\\
e^{i\frac{3\pi}{4}} & 0 & 0 & 0
\end{array}
\right).\label{C2Operator}
\end{equation}
When $\delta_x=\delta_y$, the system has an additional $C_4$ symmetry
\begin{equation}
C_{4}{h\left( k_{x},k_{y}\right)  C_{4}^{-1}=h\left(  k_{y},-k_{x}\right)}, \quad
C_{4}=\left(
\begin{array}
[c]{cccc}%
0 & 0 & e^{i\frac{5\pi}{8}} & 0\\
e^{i\frac{5\pi}{8}} & 0 & 0 & 0\\
0 & 0 & 0 & e^{i\frac{5\pi}{8}}\\
0 & e^{i\frac{\pi}{8}} & 0 & 0
\end{array}
\right).
\end{equation}

For a general flux, to write down Bloch Hamiltonian in the momentum space, it is necessary to know the size of MUC (in main text, we use the EUC to discuss the topological phases.), which is determined by (i) the dimerization pattern and (ii) the strength of the flux per plaquette $\phi$. When a rational magnetic flux $\phi= 2\pi p/q$ ($p, q$ being mutually prime integers) is applied, the MUC contains an integer number of EUCs and a flux quantum of $2\pi$. Following these conditions, we can obtain the following relations: ($i$) when $q=4n$ ($q=4n+2$), the MUC has $q (2q)$ lattice sites, and ($ii$) when $q$ is an odd number, the MUC has $4q$ lattice sites. Once determining the MUC and the gauge choice(see Fig.~\ref{fig:Gauge}), we can easily write the Hamiltonian in the momentum space.

\section{Calculation of the gapless region of the DHM at $\frac{1}{2}$ filling for a flux of $\pi/2$}
In Sec. \ref{Sec:half}, we used the Wannier-sector polarization to obtain the phase diagram at $\frac{1}{2}$ filling for a flux of $\frac{\pi}{2}$. There is a gapless region in the phase diagram that can be numerically solved. Here, we calculate analytically the boundaries of that gapless region at $\frac{1}{2}$ filling. From the Hamiltonian in Eq.~\ref{Bloch Hamiltonian4}, we can manually determine the conditions for the bulk to close its band-gap at zero energy. These are
\begin{equation}
2\delta_{x}=(1-\delta_{y}^{2})\cos(k_{y}),\nonumber\\
2\delta_{y}=(1-\delta_{x}^{2})\cos(k_{x}).
\label{gap-closing}
\end{equation}

These expressions are plotted as the red curve in Fig.~\ref{fig:Gap}a.

\section{Phases at $\frac{1}{4}$ filling for fluxes $\frac{\pi}{4}$ and $\frac{2\pi}{3}$}
\label{OtherFlux}
As mentioned in Sec. \ref{Sec:quarter}, the system can be in a trivial insulator phase, an OAL phase or a Chern insulator phase at $\frac{1}{4}$ filling for flux $\frac{\pi}{2}$. These phases can be determined by the eigenvalues of the $C_2$ symmetry operator. Here we present the phase diagrams at $\frac{1}{4}$ filling for other fluxes, and show that the OAL phases and Chern insulator phase also appear at other flux values.

\begin{figure}[ptbh]
\centering
\includegraphics[scale=0.85]{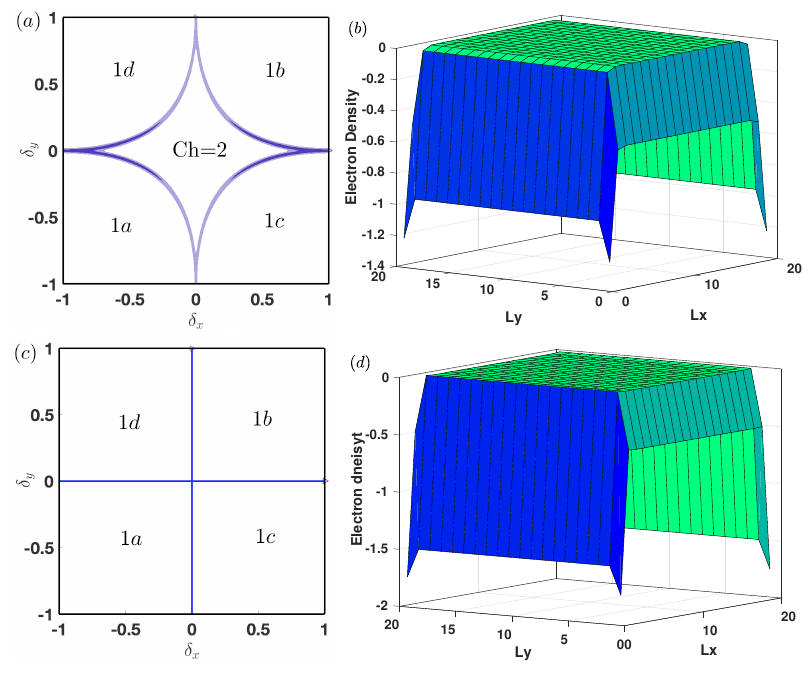}
\caption{(a) and (c) Phase diagram at $\frac{1}{4}$ filling for $\phi=\frac{\pi}{4}$ and  $\phi=\frac{2\pi}{3}$ as a function of the dimerization variables $\delta_x$ and $\delta_y$. The $1a$, $1b$, $1c$, and $1d$ stand for the Wannier center pinned to the corresponding Wyckoff position of the EUC similar to the $\frac{\pi}{4}$ case. (b) and (d) Charge density after filling the lowest $(N_x/2-1)(N_y/2-1)$ states (close to $\frac{1}{4}$-filling) under OBC with $(\delta_x,\delta_y)=(0.7,0.7)$ in $1b$ phase.The system size is $L_x=L_y=20$ MUC for the two fluxes.}
\label{fig:FigOtherFluxes}
\end{figure}

Specifically, we discuss the phase diagrams at $\frac{1}{4}$ filling for fluxes $\frac{\pi}{4}$ and $\frac{2\pi}{3}$, which are shown in Fig.\ref{fig:FigOtherFluxes}a and Fig.\ref{fig:FigOtherFluxes}c, respectively. For flux $\phi=\frac{\pi}{4}$ at $\frac{1}{4}$ filling there are five phases, just as in the case with $\phi=\frac{\pi}{2}$. In the central region, the model is in a Chern insulator phase with Chern number $C=2$. In the top right phase ($1b$ OAL phase), the charge density (subtracted background ionic charge) is shown in Fig. \ref{fig:FigOtherFluxes}b for a filling of the lowest $(N_x/2-1)(N_y/2-1)$ states (a filling of $N_xN_y/4$ states corresponds to exact $\frac{1}{4}$ filling) for a system with $N_x=2N_y=80$ lattice sites (the MUC size is $4\times 2$) under OBC. Clearly, the edge charge and corner charge deficit appear. For flux $\frac{2\pi}{3}$, the MUC has 12 ($6\times 2$) lattice sites and there are 3 electrons per MUC at $\frac{1}{4}$ filling. The three different OAL phases at $\frac{1}{4}$ filling are shown in Fig. \ref{fig:FigOtherFluxes}c.

\section{Proof of non-zero Chern number for the filling below $1/4$}
\label{Sec:Proof}
In this section, we will show that the Chern number must be non-zero for the filling smaller than $\frac{1}{4}$ for our model if the system has a gap. We will keep the discussion below as general as possible. It is important to distinguish one plaquette which only includes one lattice site, the EUC which includes dimmerization and the MUC. We assume one orbital in one plaquette and $L$ plaquettes in one EUC, so that the orbital number in one EUC is $L$. The magnetic flux in one EUC is chosen as $\phi'=\frac{2\pi p'}{q'}$, where the integers $p'$ and $q'$ are mutually prime, and thus one MUC contains $q'$ EUC, $Lq'$ plaquettes and $Lq'$ orbitals. We further assume $r$ bands in magnetic BZ are fully filled, which corresponds to $r$-orbitals in one MUC, so the filling of the system is given by $v=\frac{r}{Lq'}$. For three positive integers $r,q',p'>0$, we can always find two integers s and t to satisfy the Diophantine equation
\begin{equation}\label{eq:Diophantine}
	r=q's+p't.
\end{equation}

Next we hope to relate the Diophantine equation to the Hall conductivity $\sigma_{xy}$ through the Streda's formula. We consider the whole system contains $N$ EUCs, which correspond to $\frac{N}{q'}$ MUCs, and the 2D area is $S$. Correspondingly, each band in the magnetic BZ contains $\frac{N}{q'}$ states (different momentum values), and with $r$ bands fully filled, the total electron number in the whole system is $\frac{rN}{q'}$, and the corresponding electron density is given by $\rho= \frac{rN}{q'S}$.
Together with the Eq. (\ref{eq:Diophantine}), the electron density $\rho$ can be related to the magnetic flux by
\begin{equation}\label{eq:rho_Diophantine}
	\rho=\frac{ N s}{S}+\frac{ N t}{S}\frac{p'}{q'}=\frac{ N s}{S}+\frac{ N t}{2\pi S}\phi'
    =\frac{ N s}{S}+Bt\frac{e}{h},
\end{equation}
where $B$ is magnetic field strength. In the last step of the derivation, we have used $B=\frac{\hbar}{e}\frac{N\phi'}{S}$, which gives $\frac{p'}{q'}=\frac{BS}{N}\frac{e}{h}$ ($h=2\pi \hbar$).

On the other hand, when an integer number ($r$) of bands are fully filled, the Hall conductivity of the system should be
\begin{equation}
 \sigma_{xy}=\frac{e^2}{h}\int d_{kx} d_{ky}F_{xy}(k)=\frac{C e^2}{h}
\end{equation}
according to the TKNN formula, where the integer $C$ is the Chern number. According to the Streda's formula
\begin{equation}
 \sigma_{xy}=e\frac{\partial\rho}{\partial B},
\end{equation}
we expect the electron density $\rho$ should take the form
\begin{equation}\label{eq:rho_CN}
	\rho=A_0+BC\frac{e}{h}.
\end{equation}
By comparing Eq. (\ref{eq:rho_CN}) with Eq. (\ref{eq:rho_Diophantine}), one can see that $t=C$ should be a solution of the Diophantine equation. One should note that the Diophantine equation has infinite number of solutions since for a solution $(s,t)$, one can show ($\tilde{s}=s-kp'$, $\tilde{t}=t+kq'$) is also a solution with an arbitrary integer $k$.

Now we will address the question at which filling the Chern number can be zero. To see that, we set $t=C=0$ to be a solution of the Diophantine equation, and we should have $r=q's$, where the number $r$ of the filled bands in the MUC should be a multiply of two integers $q'$ and $s$. Correspondingly, the filling should satisfy $v=\frac{r}{Lq'}=\frac{s}{L}$ for any case with zero Chern number. For the filling $\nu<\frac{1}{L}$, the above condition cannot be satisfied and thus, the Chern number must be non-zero. This condition clearly works for the standard Hofstadter model.

In our paper, one EUC includes 4 plaquettes and thus $L=4$. The magnetic flux in one plaquette is $2\pi p/q$, so that the relation between $q,p,q'$ and $p'$ is $q'=q/4, p'=p$ when $q=4n$; $q'=q, p'=4p$ when $q=4n+1$ or $4n+3$; and $q'=q/2, p'=2p$ when $q=4n+2$. With these relations, one can check if our numerical results can satisfy the requirement of the Diophantine equation. For example, our numerical calculations show $C=1$ at the filling $v=\frac{1}{8}$ for a flux of $\pi/4$ in one plaquette. In this case, we have $q'=2,p'=1 $, $r=1$. Thus, one can show $(s,t)=(0,1)$ with $t=C=1$ is indeed a solution. We also find $C=-1$ at $\frac{1}{20}$ for a flux of $2\pi/5$ in one plaquette, and in this case $q'=5,p'=4 $, $r=1$. Thus, one can show $(s,t)=(1,-1)$ is a solution of the Diophantine equation. These examples show that our numerical calculation is consistent with the Diophantine equation.

\end{document}